\documentclass[prl,twocolumn,longbibliography,superscriptaddress]{revtex4}
\usepackage{graphicx}
\usepackage{color}
\usepackage{amssymb}
\usepackage{amsmath}

\include{NewCommands}

\usepackage{hyperref}

\definecolor{mygreen}{rgb}{0,0.5,0}
\definecolor{myred}{rgb}{0.75,0,0}
\definecolor{myblue}{rgb}{0,0,0.75}
\definecolor{mymagenta}{cmyk}{0,1,0,0.12}
\definecolor{mycyan}{cmyk}{1,0,0,0.12}
\definecolor{myorange}{rgb}{1,0.5,0}

\newcommand{\PRLsection}[1]{\noindent \textit{#1} ---}

\usepackage{amsfonts}

\begin{document}

\title{
Correlation function of spin noise due to atomic diffusion
}

\newcommand{\Princeton}
{
\affiliation{Department of Physics, Princeton University, Princeton, New Jersey 08544, USA}}

\author{V. G. Lucivero}
\Princeton
\email[Corresponding author: ]{lucivero@princeton.edu }
\author{N. D. McDonough}
\Princeton
\author{N. Dural}
\Princeton
\author{M. V. Romalis}
\Princeton

\date{\today}

\begin{abstract}
We use paramagnetic Faraday rotation to study spin noise spectrum from unpolarized Rb vapor in a tightly focused
probe beam in the presence of N$_2$ buffer gas. We derive an analytical form for the diffusion
component of the spin noise time-correlation function in a Gaussian probe beam. We also obtain analytical forms for the  frequency spectrum  of the spin noise in the limit  of a tightly focused or  a collimated Gaussian beam in the presence of diffusion. In particular, we find that in a  tightly focused probe beam the spectral lineshape can be independent of the buffer gas pressure. Experimentally, we find good agreement between the calculated and measured spin noise spectra for N$_2$ gas pressures ranging from 56 to 820 torr.

\end{abstract}

\maketitle

\newcommand{\bg}{S_{\theta}}
\newcommand{\amp}{S_{\rm A}}
\newcommand{\area}{A}
\newcommand{\gam}{\gamma}
\newcommand{\nuL}{\nu_{\rm L}}
\newcommand{\nures}{\nu_{0}}
\newcommand{\conv}{\otimes}
\newcommand{\bigconv}{\bigotimes}
\newcommand{\supin}{^{(\rm in)}}
\newcommand{\supout}{^{(\rm out)}}
\newcommand{\xysig}{\sigma}
\newcommand{\Ssig}{\rho}
\newcommand{\BinxAve}{{\cal N}}
\newcommand{\Nbin}{N_{\rm bin}}

Measurements of the quantum spin noise provide an ideal way of studying diffusion because they allow one to follow motion of identical atoms in thermal equilibrium that are distinguishable only by their intrinsic quantum fluctuations. All other techniques for studying diffusion rely on non-equilibrium spin populations or other methods of labeling atoms, for example through isotopic composition. From a practical point of view, atomic diffusion can limit the efficiency of  light-atom interaction processes, such as electromagnetically induced transparency (EIT) \cite{Xiao2006,Sargsyan2010}, coherent population trapping (CPT) \cite{Kuchina2016}, and four-wave-mixing \cite{McCormick2008}. It can also affect applications, such as quantum memories and repeaters \cite{Duan2001,Wal2003,Appel2008} and the sensitivity of atomic sensors \cite{Bloom2014} and optical magnetometers \cite{Budker2000,Kominis2003,Ricardo,Lucivero2014}.

In general the intrinsic atomic spin fluctuations, i.e. spin noise, becomes more relevant when the light-atom interaction area is reduced \cite{Crooker2004}, i.e. when the probe light is strongly focused, entering into a diffusion-limited scenario. While spin projection noise is a fundamental noise source in atomic measurements \cite{Wineland1994,Santarelli1999,Budker2007,Koschorreck2010}, it is the main source of information in spin noise spectroscopy (SNS), a powerful technique for measuring physical properties of unperturbed spin systems \cite{Pershin}, both in atomic \cite{Zapasskii2013} and solid state physics \cite{Hubner2014,Tarasenko}. Reducing the probe beam size is then desirable in SNS of atomic ensembles \cite{Zap1981,Katsoprinakis2007,Li2011,Lucivero2016}, as well as in semiconductor bulk crystals \cite{Oestreich2005,Crooker2010,Horn2013}, quantum wells \cite{Mueller2008,Poltavtsev2014} and quantum dots \cite{Glasenapp2016}. For a tightly focused probe, understanding how atomic diffusion limits atomic sensors \cite{Sheng2013} and contributes to the extracted information in SNS \cite{Mueller2008} is then an important goal with a broad range of applications.\\

A first quantitative approach for analyzing the effect of diffusion on quantum spin noise was introduced for  SNS of quantum wells \cite{Mueller2008}, in multipass-cell magnetometry \cite{Sheng2013} and in two-beam SNS \cite{Pershin2013}. However, to the best of our knowledge, neither a general analytical expression of the diffusion time-correlation function and the resonance lineshape, nor a conclusive comparison between theory and experiment has been reported. In contrast, here we derive a full analytical model for the diffusion component of the spin noise time-correlation function, based on Green's function for a general spin evolution with known relaxation rate and diffusion coefficient, which is valid over a practical range of beam waists. The theory predicts that the decay time of the diffusion correlation function follows a power law in the strong focusing case and it is not modified by an increase in buffer gas pressure.  We experimentally study the effect of diffusion on the spin noise spectra of warm Rb vapor to test the model predictions under different conditions of beam waist down to $2$ $\mu$m and buffer gas pressure up to $820$ Torr. We report good agreement of the theory with experimental spectra over all the investigated parameter range.\\

\PRLsection{Theoretical analysis}
We detect spin fluctuations of a warm unpolarized Rb vapor by paramagnetic Faraday rotation \cite{Zap1981,Crooker2004} of a far-detuned linearly polarized probe.
The spin noise power spectrum $S(f)$ can be expressed using  Wiener-Khinchin theorem as the Fourier transform of the time autocovariance of the Faraday rotation signal $\phi(t)$:
\begin{eqnarray}
S(f)&=&\int_{-\infty}^{\infty}\langle\phi(t)\phi(t+\tau)\rangle e^{-i 2\pi f \tau}d\tau \nonumber \\&=&\langle\phi(t)^2\rangle\int_{-\infty}^{\infty}C(\left\vert \tau \right\vert)e^{-i 2\pi f \tau}d\tau,
\label{eq:SpectrumTheory}
\end{eqnarray}
where $C(\tau)$ is the normalized spin time-correlation function. For each isotope the paramagnetic Faraday rotation per unit probe laser path length is given by \cite{Happer1}:
\begin{equation}
\frac{d\phi}{dl}=c r_{e}f_{osc}n \sum_{F=I\pm1/2}\frac{(\nu-\nu_F)\left\langle s_{z} \right\rangle_{F}}{(\nu-\nu_F)^2+\Gamma^2},
\end{equation}
where  $r_e$ is the classical electron radius, $f_{osc}$ is the oscillator strength of the  Rb $D1$ transition, $n$ is the number density, $\nu$ is the laser frequency and $\nu_F$ are the resonance frequencies of the two ground hyperfine states, neglecting  the excited state hyperfine splitting. The electron spin expectation  value $\left\langle s_{z}\right\rangle_{F}=\sum_m\left \langle F,m|\rho s_z|F,m\right\rangle$, where $\rho$ is the Rb density matrix.

The angle $\phi$ is measured by imaging the probe beam onto a polarizing beam splitter and  calculating normalized power difference  in the two polarizer arms, $\phi=(P_{1}-P_{2})/2(P_1+P_2).$ Therefore, the rotation angle  is given by:
\begin{equation}
\phi=\frac{\int I(\mathbf{r})(d\phi /dl)(\mathbf{r})d^{3}\mathbf{r}}{\int I(\mathbf{r})dxdy},
\end{equation}
where $I(\mathbf{r})$ is the intensity distribution of the probe beam, propagating in the $\hat z$ direction. We make measurements in  a magnetic field of  about 1 G along $\hat{y}$ direction, where the Larmor frequency of the spins far exceeds their spin-exchange relaxation rate. Under these conditions, one can ignore the correlation between  $\left\langle s_{z}\right\rangle_{I+1/2}$ and $\left\langle s_{z}\right\rangle_{I-1/2}$ states \cite{Sinutsyn}.
The time autocovariance of the rotation signal is then given by:
\begin{eqnarray}
&&\langle\phi(t)\phi(t+\tau)\rangle=\sum_{F=I \pm 1/2\ }\Big(\frac{cr_ef_{osc}nD(\nu-\nu_F)}{\int I(\mathbf{r})dxdy}\Big)^2 \times \nonumber \\
&&\int I(\mathbf{r}_1)I(\mathbf{r}_2)\left\langle s_{z}(\mathbf{r}_{1},t) s_{z}(\mathbf{r}_{2},t+\tau)\right\rangle_{F}d^3\mathbf{r}_1d^3\mathbf{r}_2, \label{eq:autocov}
\end{eqnarray}
where $D(\nu-\nu_F)=(\nu-\nu_F)/[(\nu-\nu_F)^2+\Gamma^2]$ and $\Gamma$ is the optical resonance half-width.

The evolution of the atomic density matrix is given by:
\begin{equation}
\frac {d\rho}{dt}=D \nabla^2\rho- \frac{i}{\hbar}  [H,\rho]+\mathcal{L}\rho,
\end{equation}
representing  the effects of diffusion, coherent evolution and spin relaxation. Neglecting the details of the Hamiltonian $H$ and the Lindblad relaxation superoperator $\mathcal{L}$ for multi-level Rb atoms, we represent   the atomic spin evolution by Green's diffusion function, spin precession at Larmor frequency $\omega_L$, and transverse spin relaxation time $T_2$:
\begin{eqnarray}
&&( s_{z}+is_{x})(\mathbf{r}_{2},\tau)= \\
&&\int G(\mathbf{r}_1-\mathbf{r}_2,\tau)  (s_{z}+is_{x})(\mathbf{r}_{1},0)e^{-i\omega_L \tau-\tau/T_2}d^3\mathbf{r}_1. \nonumber
\end{eqnarray}
The covariance of the spin expectation values for unpolarized atoms at two points $\mathbf{r}_1$ and $\mathbf{r}_2$  at equal times is:
\begin{eqnarray}
\left\langle s_{z}(\mathbf{r}_{1},t)s_{z}(\mathbf{r}_{2},t)\right\rangle_{F}=\left\langle s_{z}^2\right\rangle_{F} \frac{\delta(\mathbf{r}_1-\mathbf{r}_2)}{n}
\end{eqnarray}and
\begin{equation}
 \left\langle s_{z}^2\right\rangle_{F}= \frac{2F+1}{2 (2I+1)} \frac{1}{(2I+1)^2} \frac{F(F+1)}{3},
 \end{equation}
where the first ratio gives the fraction of atoms in state $F$, the second  is the equal to $(s_z/F_z)^2$ and the last is $\left\langle F_{z}^2\right\rangle$ in unpolarized $F$ state. The spin covariance function entering Eq. \ref{eq:autocov} is then given by:
 \begin{eqnarray}
&&\left\langle s_{z}(\mathbf{r}_{1},t)s_{z}(\mathbf{r}_{2},t+\tau)\right\rangle_{F}= \nonumber \\
&&\left\langle s_{z}^2\right\rangle_{F} G(\mathbf{r}_1-\mathbf{r}_2,\tau)\cos(\omega_L \tau)e^{ -\tau/T_2}/n,
\end{eqnarray}
where the diffusion Green's function is given by $G(r,\tau)=e^{-r^2/4D\tau}/(4\pi D\tau)^{3/2}$.

Since the Green's function approaches the Dirac delta function at $\tau=0$, the total rotation angle integrated noise power, in units of rad$^2$, can be written as:
\begin{equation}
\langle\phi(t)^2\rangle=\sum_{F=I \pm 1/2} (cr_ef_{osc}nlD(\nu-\nu_F))^2\frac{\left\langle s_{z}^2\right\rangle_F}{N_{eff}},
\label{eq:variance}\end{equation}
where
\begin{equation}
N_{eff}=nV_{eff}=n\frac{(\int I(\mathbf{r})d^3\mathbf{r})^2}{\int I(\mathbf{r})^2d^3\mathbf{r}} \end{equation}
and $l$ is the cell length. This equation can be interpreted as giving optical rotation from a sample of length $l$ with spin polarization given by $\sqrt{\left\langle s_{z}^2\right\rangle_F/N_{eff}}$.  The r.m.s.~fluctuations in the spin polarization are proportional to inverse square root of the total effective number of atoms participating in the measurement, as expected.

The normalized diffusion correlation function is then given by:
\begin{equation}
C_{d}(\tau)=\frac{\int I(\mathbf{r}_1) G(\mathbf{r}_1-\mathbf{r}_2,\tau)I(\mathbf{r}_2) d^3\mathbf{r}_1d^3\mathbf{r}_2}{\int I(\mathbf{r})^2d^3\mathbf{r}}.
\label{eq:cd}\end{equation}

This analysis applies to an arbitrary intensity distributions, including multi-pass cells \cite{Sheng2013}. Here we focus on a simple case of a single Gaussian probe  beam with a waist radius $w_0$ at the center, $z=0$, of a single pass cell of length $l$. The Gaussian beam has a Rayleigh range $z_R=\pi w_0^2/\lambda$, where $\lambda$ is the probe beam wavelength. By assuming that the cell radial dimension is much bigger than the maximum beam radius, starting from Eq. (\ref{eq:cd}), one can write the diffusion  spin noise time-correlation function as:
\begin{equation}
C_d(\tau)=\int\limits^{\frac{l}{2}}_{-\frac{l}{2}}\int\limits^{+\frac{l}{2}}_{-\frac{l}{2}}\frac{V_{eff}}{l^2\sqrt{\pi^3D \tau}}\frac{   e^{-\frac{(z_1-z_2)^2}{4D\tau}}dz_1dz_2}{w(z_1)^2+w(z_2)^2+8D\tau},
\label{eq:Modelz1z2}
\end{equation}
where $w(z)$ is the beam radius at position $z$,  the integrals run over the probe propagation length of the cell $l$, and the effective  beam volume is $V_{eff}=\lambda l^2/{2\arctan(  l/2 z_{R})}$.   The evaluation of the integral in Eq. (\ref{eq:Modelz1z2}) can be simplified by using an asymptotic expansion of the Error function for large $x$, Erfc$(x)\approx e^{-x^2}/\sqrt{\pi}x$, which is generally quite accurate because $x\approx 2\pi w_0/\lambda=2/(NA)$, where $NA$ is the numerical aperture of the Gaussian beam.  The fractional correction to the leading order expansion is on the order of  $ (NA)^2$/16, much less than one, unless the probe beam is very tightly focused.  Under these conditions we obtain an analytical solution:
\begin{equation}
C_d(\tau)=\frac{w_0 \arctan^{}{\frac{lw_0}{2z_R \sqrt{4\tau D+w_0^2}}}}{\sqrt{4\tau D+w_0^2}\arctan{(l/2z_R)}}.
\label{eq:Model}
\end{equation}.

The correlation function derived in Eq.~(\ref{eq:Model}) can be further simplified in two limits, when the probe beam is nearly collimated over the length of the cell, $z_R\gg l$, and when the beam is tightly focused and quickly diverges inside the cell, so $z_R\ll l$.

For a well-collimated beam inside the cell with $z_R\gg l$ we get
\begin{equation}
C_d(\tau)=\frac{1}{1+4\tau D/w_0^2}, \label{Eq:Cdcoll}
\end{equation}
while for a tightly focused beam with $z_R\ll l$ and $4\pi w_0\sqrt{\tau D} \ll \lambda l$ we get
\begin{equation}
C_d(\tau)=\frac{1}{\sqrt{1+4\tau D/w_0^2}}. \label{Eq:Cdfocus}
\end{equation}

For these simple limiting forms of the diffusion correlation function one can perform analytically the Fourier transform in Eq.~(\ref{eq:SpectrumTheory}) to obtain the frequency shape of the probe polarization rotation noise spectrum.
For the case of the collimated laser beam, Eq.(\ref{Eq:Cdcoll}), we obtain:
\begin{eqnarray}
S(f)&=&\frac{\langle\phi(t)^2\rangle w_0^2}{2D}{\rm Re}[e^sE_1(s)], \label{Eq:Spcoll}\\
s&=&\frac{w_0^2}{4 D}\left[1/T_2+2\pi i(f-f_L)\right],
\end{eqnarray}
where $E_1(s)=\int_{s}^{\infty}(e^{-x}/x) dx$ is the exponential integral. Here $T_2$ is the intrinsic atomic transverse spin relaxation time, neglecting relaxation due to probe beam scattering, and $f_L=\omega_L/2\pi$ is the Larmor frequency.  When $w_0^2/4D T_2 \gg 1 $ the diffusion time across the beam is much longer than the intrinsic spin coherence time, and the lineshape reduces to a Lorentzian with a width equal to $T_2$.
Eq.~(\ref{Eq:Spcoll}) allows one to calculate an analytical form for the spectrum in a common case when diffusion somewhat broadens the spin noise lineshape. It is similar to the spectrum of probe beam absorption fluctuations obtained for a collimated Gaussian beam in \cite{Aoki}.

In the other limit when the probe beam is tightly focused in a cell that is much longer than the Rayleigh range,  the correlation function is given by Eq. (\ref{Eq:Cdfocus}) and we obtain a lineshape:
\begin{equation}
S(f)=\frac{\langle\phi(t)^2\rangle w_0 }{\sqrt{  D}}{\rm Re}\left[\frac{\sqrt{\pi}}{\sqrt{1/T_2+2 \pi i (f-f_L)}} \right], \label{Eq:Spfocused}
\end{equation}
which is valid for $w_0/\sqrt{D T_2} \ll 1$ and for $w_0\sqrt{2 \pi (f-f_L)/D} \ll 1$. This limit corresponds to the diffusion time across the beam waist being much faster than the spin coherence time or the inverse of the frequency detuning.

An interesting aspect of the limit in Eq.~(\ref{Eq:Spfocused}) is that the lineshape  is independent of the diffusion constant and therefore is independent of the buffer gas pressure. This limit only applies if Eq.~(\ref{Eq:Cdfocus}) is valid, therefore only if $4\pi w_0\sqrt{\tau D} \ll \lambda l$, which  can be rewritten as $w_e^2/4\tau D\gg1$, where $w_e$ is the beam radius at the ends of the cell. Thus, if the laser beam is tightly focused such that the diffusion timescale is much faster than the relaxation timescale at the waist of the  beam, $w_0^{2}/4D T_2 \ll 1,$ and at the same time  it  is  much slower than the relaxation time scale at the ends of the cell, $w_e^2/4DT_2\gg1$, we enter into a new scaling regime.   In this regime, the lineshape is roughly a fourth root of a Lorentzian  with a  half-width that  depends only on $T_2$. For a certain range of parameters  the lineshape for a focused beam can be even narrower than for a collimated beam with $z_R\gg l$.

\PRLsection{Numerical analysis} Our analysis assumes that the mean free path of Rb atoms is much smaller than the smallest beam waist size, so the motion of the atoms remains in the diffusion regime. The atom mean free path $l_f=1/\sigma n_{\text{N}_2}$, where $\sigma$ is the cross-section for velocity-changing collisions, which is on the order of $5\times 10^{-15}$~cm$^2$ \cite{Gallagher}. At  N$_2$ pressure of 50 torr we get  $l_f=1~\mu$m, so  for $w_0=2~\mu m$ and N$_2$ pressure greater  or equal to about 50 torr the atom motion is in the diffusion regime.

We use the diffusion constant for Rb-N$_2$ that was measured in \cite{Ishikawa2000},
$D_0=0.159$ cm$^2$/s at 60$^{\circ}$C. The diffusion constant scales with temperature and pressure as $D(p_{N_2})=D_0 (p_0 /p_{N_2})(T/T_0)^{3/2}$, where $p_0=760$ Torr and $T_0$ is the absolute temperature at which $D_0$ is measured.

\begin{figure}[t]
\centering
  \hspace{0.4cm}
\includegraphics[width=\columnwidth]{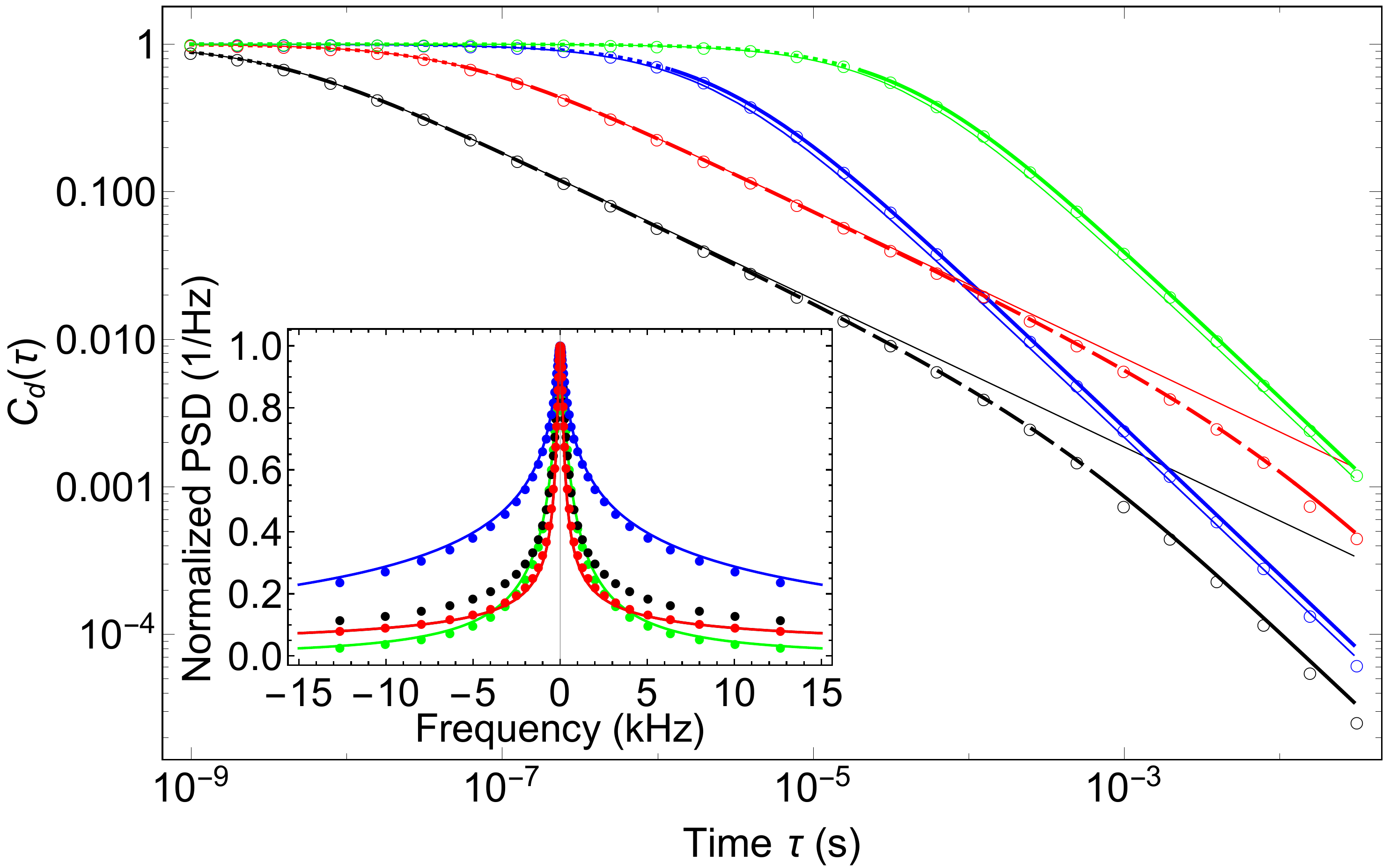}
 \caption{\textbf{Calculated diffusion time-correlation function:}
   The points are numerical results  from Eq. (\ref{eq:Modelz1z2})  and thick lines are calculated from Eq.~(\ref{eq:Model}) for   $w_0^{(1)}=2$ $\mu$m and $p_{N_2}=50$ Torr (black)  and $p_{N_2}=800$ Torr (red); for $w_0^{(2)}=50$ $\mu$m and  $p_{N_2}=50$ Torr (blue)  and $p_{N_2}=800$ Torr (green). Simplified analytical results are shown by thin lines from Eq.~(\ref{Eq:Cdfocus}) (black and red) for  $w_0^{(1)}=2~\mu$m and from Eq.~(\ref{Eq:Cdcoll}) (blue and green) for  $w_0^{(2)}=50~\mu$m. Dotted, dashed and solid lines indicate transitions between different diffusion regimes.
\textbf{Inset: Calculated spectra.} Normalized spin noise power spectral densities using the same colors for $T_2=1$ msec. Solid lines show analytic spectra, Eq.~(\ref{Eq:Spfocused}) and Eq.~(\ref{Eq:Spcoll}) for  $w_0^{(1)}=2$ $\mu$m and  $w_0^{(2)}=50~\mu $m respectively. Points show numerical calculation  using  Fourier transform of Eq. (\ref{eq:Model}).}
  \label{fig:Theory}
\end{figure}

In Fig. (\ref{fig:Theory}) we show the theoretical diffusion component of the spin time-correlation function, calculated numerically from Eq. (\ref{eq:Modelz1z2}) (points) and analytically from Eq. (\ref{eq:Model}) (thick lines) for a beam waist of $w_0^{(1)}=2$ $\mu$m and $w_0^{(2)}=50$ $\mu$m at buffer gas pressures of 50 torr and 800 torr, see caption for colors. In these simulations we take the cell length $l=15$ mm,  $\lambda=795$ nm, and $T=100^{\circ}$C, corresponding to the experimental conditions. Thin solid lines show the approximation for a collimated beam, Eq.~(\ref{Eq:Cdcoll}), for $w_0^{(2)}=50~\mu$m and the approximation for a focused beam, Eq.~(\ref{Eq:Cdfocus}), for $w_0^{(1)}=2~\mu$m.
Dotted  lines indicate the regime $\tau< w_0^2/4D$, dashed lines indicate the regime $w_0^2/4D<\tau< w_e^2/4D$, and solid thick lines indicate $\tau> w_e^2/4D$. Thus, for a focused beam there is a large range of $\tau$ where the correlation function follows a simple scaling relationship, $C_d(\tau) \propto \tau^{-1/2}$.

In the inset of Fig.~\ref{fig:Theory} we show the spin noise power spectral densities calculated for different cases using numerical Fourier transform of the correlation function, as well as the analytical limiting cases given by Eq. (\ref{Eq:Spcoll}, \ref{Eq:Spfocused}). The spectra are normalized to their peak value. One can see that for the tightly focused case the linewidth does not change significantly with the buffer gas pressure over more than an order of magnitude in pressures. The noise peak is also narrower for the tightly focused probe beam than for the collimated beam at 50 torr.  This somewhat surprising  result  is due to fast divergence of the probe beam. Near the ends of the cell the beam is wide enough that the diffusion time across the beam is slow. For the range of times $\tau$  that
correspond to the intrinsic spin relaxation time $T_2$ the correlation function is given by a power law with no characteristic time scale. Hence, the lineshape does not change with pressure of the buffer gas.

\begin{figure}[t]
\centering
  \hspace{0.4cm}
\includegraphics[width=\columnwidth]{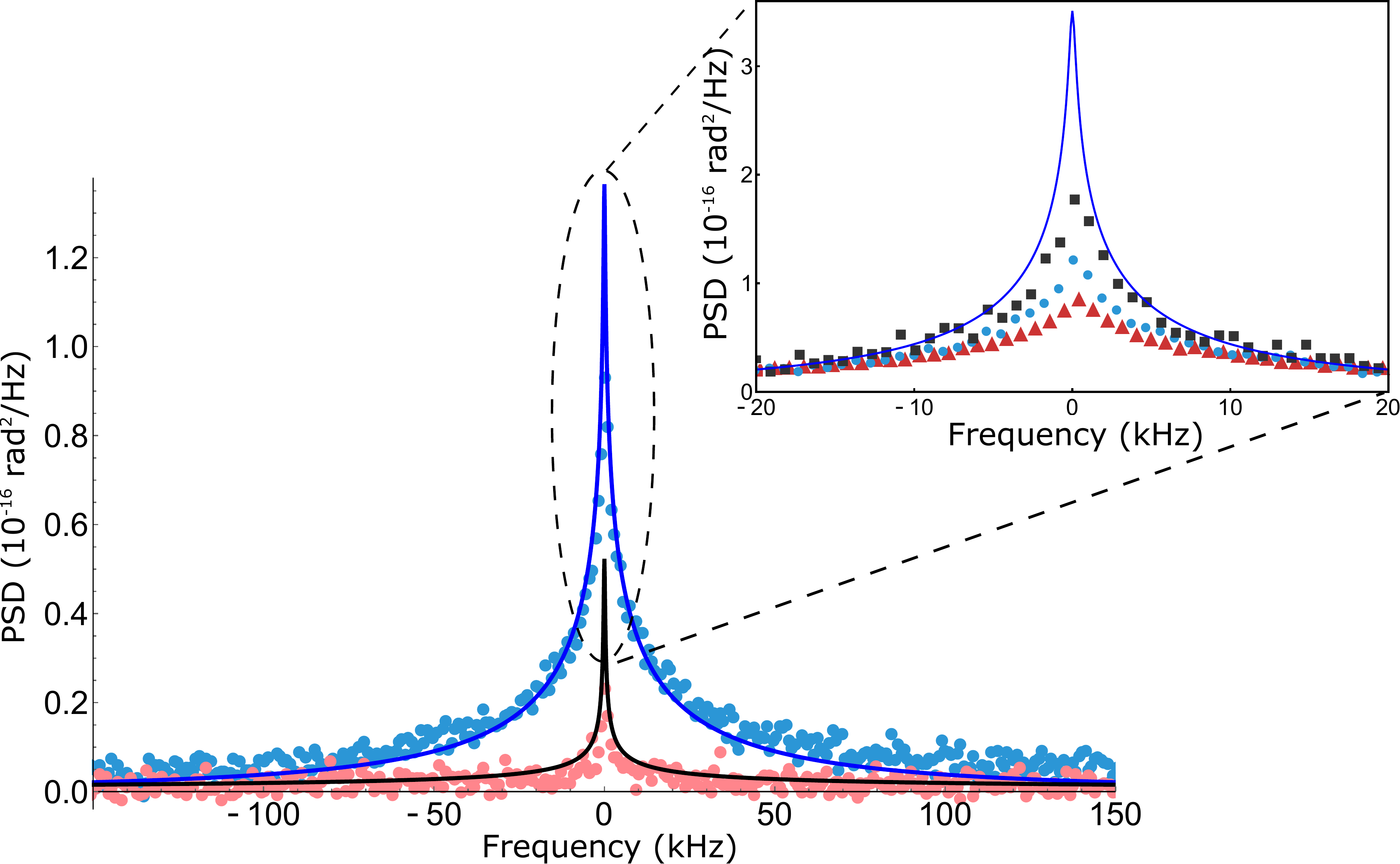}
  \caption{\textbf{Spin noise spectra.} Experimental noise spectra ($5000$ averages) for strongly focused (red points) and pseudo-collimated (blue points) probe beam. Data acquired at $T=80^{\circ}C$, detuning $\Delta=50$GHz, buffer gas pressure $p_{N_2}=56.5$ Torr and optical power $P=700\mu$W. Continuous lines are the calculated spectra for focused (black) and collimated (blue) probe, respectively.
  \textbf{Inset: Power broadening.} Experimental spectra acquired at buffer gas pressure $p_{N_2}=200$ Torr for a collimated probe ($w_0=50\mu$m) with optical power $P=300\mu$W (black squares), $P=700\mu$W (blue points) and $P=1.5 m$W (red triangles). The blue line is the calculated spectrum.}
  \label{fig:Spectra}
\end{figure}

\PRLsection{Experimental procedure}
\begin{figure}[t]
\centering
  \hspace{0.4cm}
\includegraphics[width=\columnwidth]{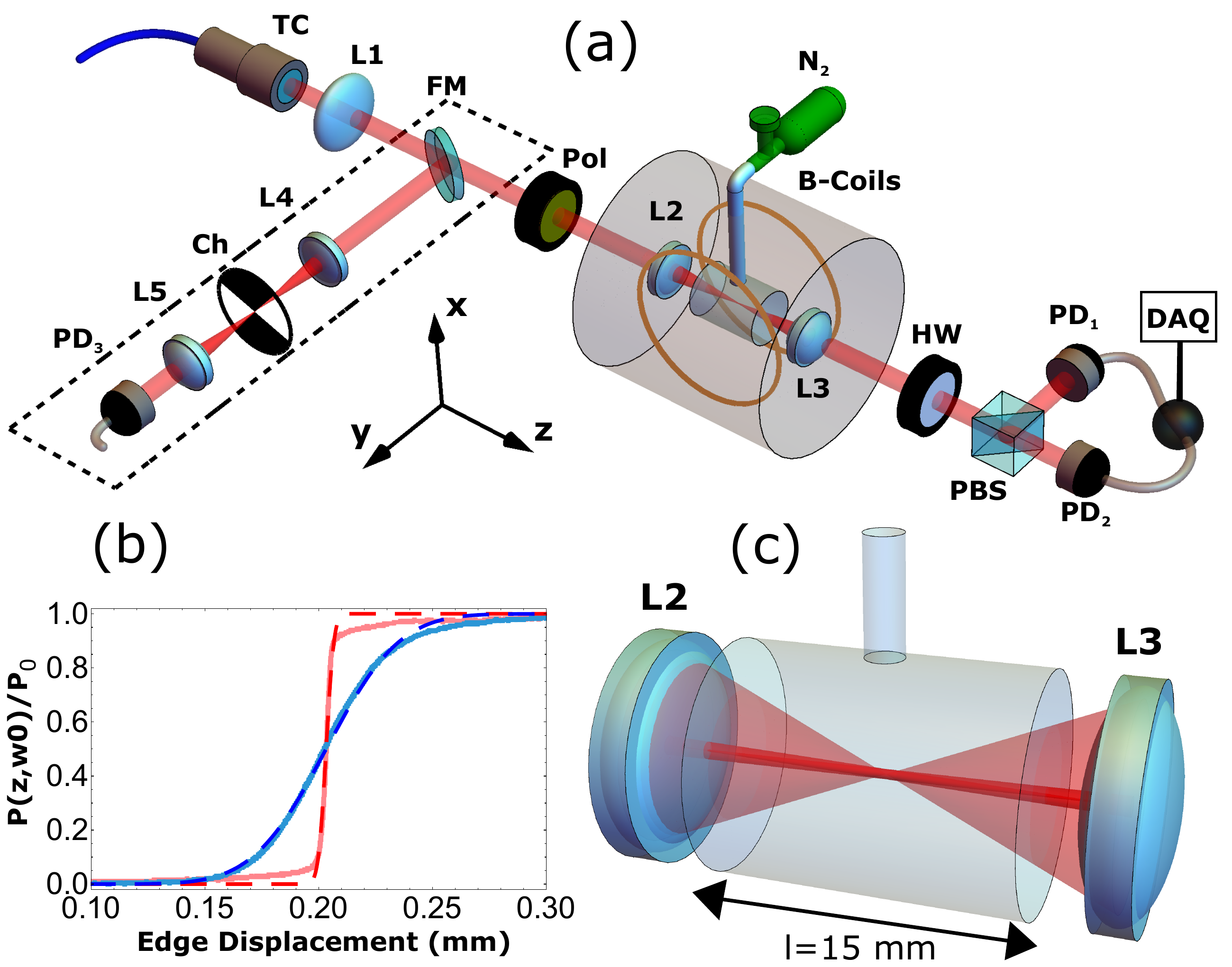}
  \caption{\textbf{a) Experimental Setup.} TC - Triplet collimator; L1 - Lens with focal f$_1=500$ mm; Pol - Linear Polarizer, L2/L3 - Lenses with focal f$_2=$f$_3=10$ mm; N$_2$ - Reservoir of nitrogen buffer gas; HW - Half waveplate; PBS - Polarizing beam splitter; PD - Photo-detector; DAQ - data acquisition card; FM - Flip mirror; L4/L5 - Lenses with focal f$_4=$f$_5=10$ mm; Ch - Chopper. \textbf{b) Waist measurement.} Normalized transmitted power  $P(w_0,z)/P_0$ versus chopper displacement for strongly focused (red) and collimated (blue) probe conditions. The dashed lines are the fit to the complementary error function.
\textbf{c) Vapor cell geometry.} Beam configurations across the vapor cell. The probe beam is either strongly focused (w$^{(1)}_0=2$ $\mu$m) or collimated (w$^{(2)}_0=50$ $\mu$m) at the center of the cell by the lens L2. The two probe conditions are obtained by removing or placing the lens L1 from the setup, respectively.}
  \label{fig:SETUP}
\end{figure}
In order to test the described model and theoretical predictions we built the experimental setup shown in Fig. (\ref{fig:SETUP} (a)). The output beam of the laser source is fiber coupled and connected to a triplet collimator, which provides a high-quality collimated beam with a Gaussian diameter of $2w=3.8$ mm. This laser is used to probe a natural abundance Rb vapor placed within a cylindrical vapor cell, made out of Pyrex, with length $l=15$ mm, diameter $d_{\text{cell}}=1$ cm with double-sided AR-coated windows anodically bonded to the ends of the cell. The cell is evacuated, baked, and filled with Rb metal using a vacuum system and then connected to a cylinder filled with about $1000$ Torr of N$_2$ buffer gas. A system of valves allows us to control the released amount of buffer gas pressure after each set of measurements. The vapor cell is placed within a boron-nitride oven, which is heated by ac current flow in twisted heating wires. The temperature is monitored by a thermocouple and stabilized to $0.1^{\circ}$C with an analog temperature controller. The entire system is enclosed in  $2$ $\mu$-metal and $1$ aluminium layers of magnetic shielding, while magnetic coils generate a field $B_y$, transverse to the probe propagation direction. As shown in Fig. (\ref{fig:SETUP} (c)) we study two different probe beam shaping configurations: a strong focused case, obtained with a molded aspheric lens (L$2$) of focal length f$_2=10$ mm, which focuses the probe to a waist radius of $w^{(1)}_0=2$ $\mu$m at the center of the vapor cell and a pseudo-collimated case, obtained by adding a plano-convex lens (L$1$) with f$_1=500$ mm at the focal distance before the first aspheric lens (L$2$), so that the probe stays approximately collimated with $w^{(2)}_0=50$ $\mu$m across the vapor cell. In both conditions, a second aspheric lens (L$3$) with f$_3=10$ mm collimates the probe beam after atomic interaction.
We measure the beam waist by using a copy of the optical system outside the shielding, which consists of two aspheric lenses L$4$ and L$5$ with focal f$_4=$f$_5=10$ mm, and a rotating optical chopper (as shown in the dashed region of Fig. (\ref{fig:SETUP} (a)). The probe beam is linearly polarized in the $x-y$ plane before atomic interaction and propagates in the $z$ axis. By applying a transverse magnetic field $B_y=0.71$ G, the intrinsic spin noise fluctuations oscillate at the Larmor frequency $\omega_L = g_F \mu_0 B_y$, where $g_F$ is the Land\'{e} factor and $\mu_0$ is the Bohr magneton. The probe beam undergoes paramagnetic Faraday rotation, which is detected by a conventional balanced polarimeter. The differential output signal is fed into a data acquisition card from which we compute the power spectral density, resulting in a spin noise spectrum, as shown in several prior works \cite{Zap1981,Crooker2004,Oestreich2005}. In order to compare the experimental spectra against theory, we measure the number density $n$ and the buffer gas pressure $p_{\text{N}_2}$ by fitting absorption spectra acquired at low probe power of $\simeq1\mu$W, as described in \cite{Romalis1997}. At fixed density, we independently measure the transverse relaxation time $T_2$ by fitting the exponential decay of low light-induced atomic polarization, generated by the probe itself, temporarily circularly polarized \cite{Li2011}.\\
\PRLsection{Results and discussion}
In Fig. (\ref{fig:Spectra}) we show the portion of the experimental spin noise spectra around the $^{85}$Rb resonance, after subtracting the photon shot noise background and shifting the peak to zero frequency, for the probe beam strongly focused at $w^{(1)}_0=2$ $\mu$m (red) and collimated at $w_0^{(2)}=50$ $\mu$m (blue), and we compare them against calculated spectra. While the total integrated noise is larger for the strongly focused beam, the fast transit time extends the lineshape wings to several MHz relative to the collimated case and the tails of the two spectral shapes cross far from resonance. Thus the collimated case with $w_0^{(2)}=50$ $\mu$m gives a greater SNR for same parameter conditions. As shown in Fig. (\ref{fig:Spectra}) for measured density $n=1.25\times10^{12}$ cm$^{-3}$, pressure $p_{N_2}=56.5$ Torr, relaxation time $T_2=1$ msec and detuning from the Rb D$1$ line $\Delta=50$ GHz, we found good agreement between the experimental spectra and the noise spectra calculated from Eq. (\ref{eq:SpectrumTheory}) and Eq. (\ref{eq:Model}) without any free parameters. While the theoretical lineshape in the wings agrees well with the data, we observe a discrepancy at the resonance peak-noise. This is mostly due to power broadening because of the residual optical pumping induced by the probe beam. Indeed, the experimental spectrum converges to the theoretical one either by reducing the probe power, as shown in the inset of Fig. (\ref{fig:Spectra}) at $p_{N_2}=200$ Torr, or increasing the optical detuning (not shown).\\
\begin{figure}[t]
\centering
  \hspace{0.4cm}
\includegraphics[width=\columnwidth]{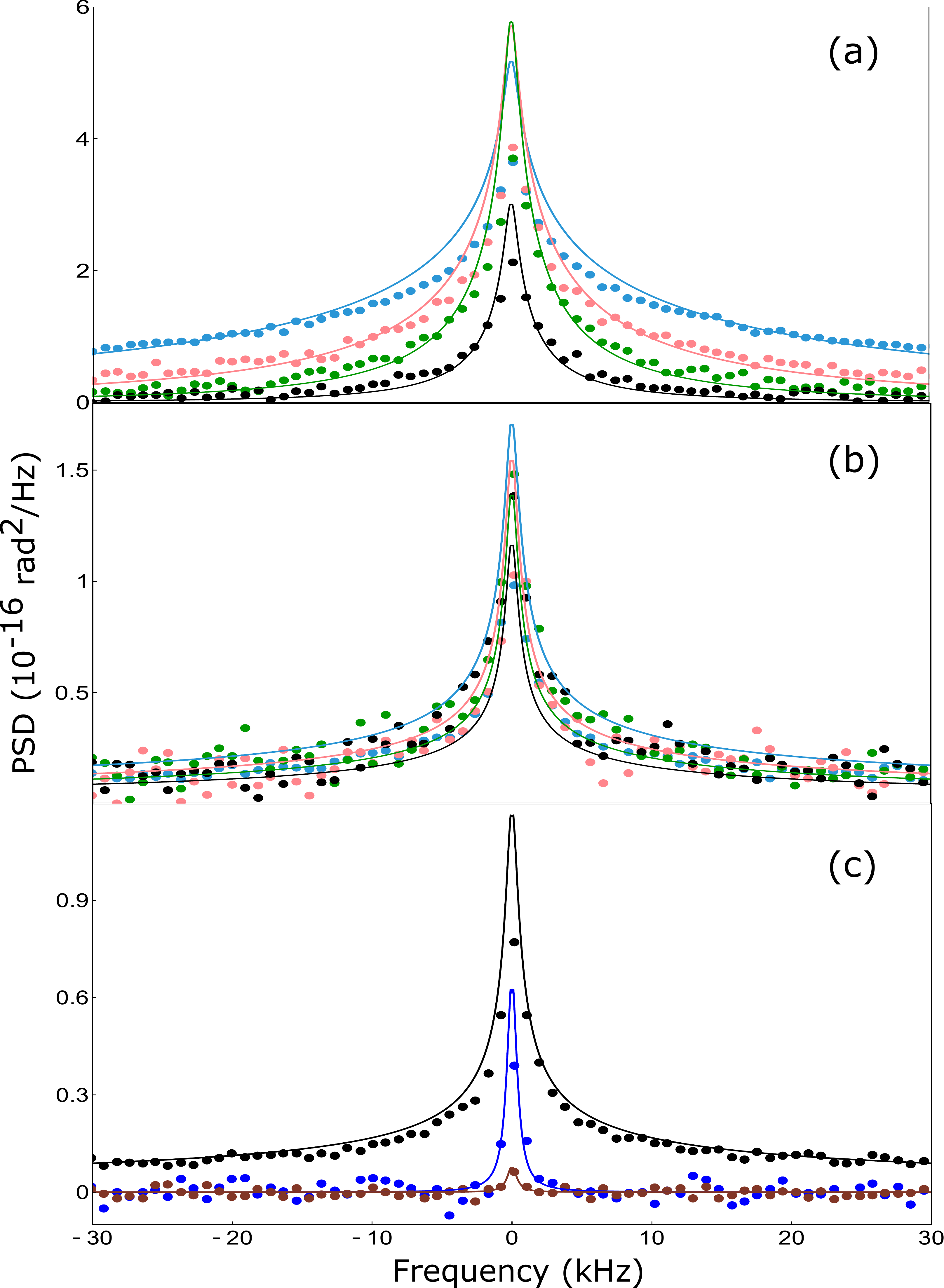}
  \caption{\textbf{Dependence on buffer gas pressure.  (a)} experimental (points) and calculated (continuous lines) spin noise spectra acquired with a pseudo-collimated probe at $T=100$ $^{\circ}C$ for conditions (from top to bottom): ($\Delta=50$ GHz, $p_{N_2}\simeq56.5$ Torr) (blue), ($\Delta=75$ GHz, $p_{N_2}\simeq200$ Torr) (red), ($\Delta=100$ GHz, $p_{N_2}\simeq500$ Torr) (green), ($\Delta=150$ GHz, $p_{N_2}\simeq820$ Torr) (black).  \textbf{(b)} experimental (points) and calculated (continuous lines) spin noise spectra acquired with a strongly focused probe at $T=100$ $^{\circ}C$ for conditions: ($\Delta=50$ GHz, $p_{N_2}\simeq56.5$ Torr) (blue), ($\Delta=75$ GHz, $p_{N_2}\simeq200$ Torr) (red), ($\Delta=75$ GHz, $p_{N_2}\simeq500$ Torr) (green), ($\Delta=75$ GHz, $p_{N_2}\simeq820$ Torr) (black). \textbf{(c)} experimental (points) and calculated (continous lines) noise spectra acquired at $T=100$ $^{\circ}C$ with $p_{N_2}\simeq820$ Torr for probe beam focused to $w_0^{(1)}=2$ $\mu$m (black), or collimated with radius $w^{(1)}=0.5$ mm (blue) and $w^{(2)}=1.5$ mm (brown). Optical probe power is $P=300\mu$W.}
  \label{fig:BufferGasDep}
\end{figure}
In Fig. (\ref{fig:BufferGasDep}) we report the study of the spin noise lineshape for different $N_2$ pressure and probe waist at higher temperature $T=100$ $^{\circ}C$, corresponding to measured density $n=6.12\times10^{12}$ cm$^{-3}$ and $T_2=0.6$ msec. At low buffer gas pressure the FWHM resonance linewidth is actually smaller in the strongly focused case, when compared with the collimated case that gives same SNR. Then, we increased the buffer gas pressure $p_{N_2}$ from $56.5$ to $820$ Torr and we found that the resonance linewidth is reduced when the probe is pseudo-collimated to $w_0^{(2)}$, while no change occurs in the strongly focused case with $w_0^{(1)}$. These effects are clearly shown in Figs. (\ref{fig:BufferGasDep}) (a) and (b) for the two beam waist conditions and are due to the different dependence of the diffusion correlation decay time on buffer gas pressure, as predicted by the theory and shown in Fig. (\ref{fig:Theory}). When we increase the buffer gas pressure and broaden the absorption cross section, we also increase the optical detuning in order to reduce optical pumping by the probe beam. While in the pseudo-collimated case we acquired data up to $\Delta=150$ GHz at the upper investigated pressure of $820$ Torr, in the strongly focused case we limited the detuning to $\Delta=75$ GHz in order to get an appreciable SNR (see captions of Figs. (\ref{fig:BufferGasDep}) (a-b)). We found good agreement between experimental and calculated spectra, i.e. our measurements validate the predicted dependence of the diffusion correlation function on the buffer gas pressure without any free parameters. As in Fig. (\ref{fig:Spectra}) the peak-noise discrepancy is due to residual optical pumping by the probe beam and is reduced for low power and large detuning.\\
For completeness, in Fig. (\ref{fig:BufferGasDep}) (c), we compare the spin noise spectrum obtained with strong focusing $w_0^{(1)}=2$ $\mu$m with two new conditions in which the probe is purely collimated with beam radius $w^{(1)}=0.5$ mm and $w^{(2)}=1.5$ mm (see figure caption), obtained by removing the lenses ($L_1,L_2,L_3$) from the experimental setup described in Fig. (\ref{fig:SETUP}) and appropriate beam shaping. Both theory and data, shown at maximum buffer gas pressure of $820$ Torr and detuning $\Delta=75$ GHz, confirm that decreasing the probe beam radius results into a larger total noise variance, justifying the focusing strategy to significantly improve the SNR. However, the transit time broadening  due to diffusion results in SNR reduction, as one can see comparing the spectra with $w_0=50$ $\mu$m and $w_0=2$ $\mu$m shown in Fig. (\ref{fig:Spectra}) and Figs. (\ref{fig:BufferGasDep}) (a-b), setting an experimental limitation to the SNS sensitivity \cite{Lucivero2017}.

\PRLsection{Conclusions} In conclusion, we described an analytical model for the diffusion component of the spin noise time-correlation function. The model is valid for varying Gaussian beam profiles, spin relaxation rates and diffusion coefficients. We derived analytical lineshapes for the spin noise spectrum in the case of a collimated or a tightly focused Gaussian probe beam. In the intermediate regime the spectral profile can be obtained by Fourier transform of the time-correlation function. We found a number of interesting features for the case of a tightly focused  Gaussian probe beam with a  Rayleigh range much smaller than the cell length. In this case the spin noise spectral lineshape can be narrower than for a collimated probe beam and it does not depend on the buffer gas pressure. A Gaussian beam with a short Rayleigh range relative to the cell length has a distribution of different diffusion time scales, which add up to give a power law time correlation function without a characteristic time scale. As a result, the lineshape is determined by the intrinsic spin relaxation time even in the presence of diffusion.

We experimentally studied the effects of atomic diffusion on the spin noise spectra of a warm Rb ensemble for different beam waists and buffer gas pressures. We found agreement with theoretical predictions without any free parameters. In particular, we confirmed that the resonance linewidth for a tightly focused probe is narrower than the one in the collimated case at equal SNR. However, transit time broadening limits the absolute SNR in the strong focusing condition, showing an experimental trade-off with the improvement given by reducing the beam area. We also found that increasing the buffer gas pressure reduces the spin noise linewidth for a collimated probe but does not change the spectral lineshape in a tightly focused beam. This work can directly improve the design of multipass cells atomic measurements \cite{Li2011,Sheng2013} and the sensitivity of SNS of warm atomic ensembles \cite{Katsoprinakis2007,Lucivero2017}. The described model can be generalized to SNS of semiconductors as bulk crystals \cite{Oestreich2005,Crooker2010,Horn2013} and nanostructures \cite{Mueller2008,Poltavtsev2014,Glasenapp2016}, where strong focusing is desiderable because of the increased spin noise power and small sample volume. In general, it can be used to improve control and optimization of spin-based atomic quantum metrology \cite{Vasilakis2015,Hosten2016,Colangelo2017}, quantum information protocols \cite{Warner2013,Krauter2013,Guendofmmodereveglsegian2015} and spintronics \cite{Wolf2001,Freeman2012}.

\PRLsection{Acknowledgements}
This work was supported by Israel MOD research grant.  We thank Attaallah Almasi, Mark Limes and Giorgio Colangelo for useful discussions.

\bibliographystyle{apsrev4-1}
\bibliography{DIFFinSNSS}
\end{document}